\documentclass[12pt,preprint]{aastex}
\pdfoutput=1
\newcommand{\As}{\rm \AA}
\newcommand{\IS}{ \bf I(\nu^{\infty}, \bf r,\Omega) }
\newcommand{\Ss}{\bf S (\nu^{\infty}, \bf r, \Omega) }
\usepackage{graphicx}

\begin{document}
\author{A. Dorodnitsyn\altaffilmark{1}, T. Kallman\altaffilmark{1}}
\altaffiltext{1}{Laboratory for High Energy Astrophysics, NASA Goddard Space Flight Center, Code 662, Greenbelt, MD, 20771, USA}

\title{Active galaxy unification in the era of X-ray polarimetry}

\begin{abstract}
Active Galactic Nuclei (AGN), Seyfert galaxies and quasars, are powered by luminous accretion and often accompanied by winds which are powerful enough to affect the AGN mass budget,
and whose observational appearance bears an imprint of processes which are happening within the central parsec
around the black hole (BH).
One example of such a wind is the partially ionized gas responsible for X-ray and UV absorption 
('warm absorbers').  
Here we show that  such gas will have a distinct signature  when viewed in  polarized X-rays.
Observations of such polarization can test models for the geometry of the flow, and the gas responsible for 
launching and collimating it.  
We present calculations which  show that the polarization depends on the hydrodynamics of the flow, 
the quantum mechanics of resonance line scattering and the transfer of  polarized X-ray light 
in the highly ionized moving gas. The results emphasize the three 
dimensional nature of the wind for modeling spectra.  We show that the polarization in the 
0.1-10 keV energy range is dominated by the effects of resonance lines.  We 
predict a $5-25\%$ X-ray polarization signature of type-2 objects in this energy range.
These results are general to flows which originate from a cold torus-like structure, located $\sim 1$pc 
from the BH, which wraps the BH and is ultimately responsible for the 
apparent dichotomy between type 1 and type 2 AGNs.
Such signals will be detectable by future dedicated X-ray polarimetry space missions, such as the 
NASA Gravity and Extreme Magnetism SMEX, GEMS \citep{Swank}.
\end{abstract}

\section{}
Optical spectropolarimetry  of Seyfert 2 galaxies 
was responsible for the revelation that both Seyfert 1 and Seyfert 2 AGNs are intrinsically the same 
objects, and that their apparent differences 
are due to the inclination from which they are viewed \citep{AntonucciMiller85}.
A paradigm of a cold ($T=100-1000$K and several parsecs in diameter) molecular torus 
which is obscuring the broad line region in most low and intermediate luminosity 
AGNs, is supported by the polarized optical line spectra  
and in some cases by direct interferometric observations \citep{JaffeNATUR}.

The obscuring torus may represent part of a flow originating at smaller
 \citep{Elit06} or larger  \citep{Proga07} radii. Studies of the lines blueshifts combined with variability studies of warm absorber spectra 
\citep{Behar03,Netzer03}, suggest a location of the torus at the distance $\gtrsim 1$pc from the BH.

First principal arguments \citep{KrolikBegelman86,KrolikBegelman88}
support the idea that the torus is the source of an outflow
which is evaporated by X-ray heating, and such gas is a plausible source of 
warm absorbers \citep{KrolikKriss1,KrolikKriss2}.

The question of the origin of the warm absorber flow is intrinsically related to the problem of the vertical
support of the torus against collapse.
Infrared radiation pressure is a likely mechanism for such support
\citep{Krolik07}, although magnetic fields could also contribute \citep{Lovelace98}.
One may also expect large scale magnetic fields playing a role in producing hydromagnetic 
flows related to warm absorbers. Models based on self-similar solutions of the type developed by 
\cite{BlandfordPayne82} were calculated by \cite{Emmering91}, \cite{Bottorff97}, and \cite{Bottorff00};
warm absorber spectra in transmission from a different type of 2D self-similar solutions of the Grad-Shafranov equation
were calculated by \cite{Fukumura09}.

Warm absorber gas is heated and ionized by X-rays from the BH and in the low density limit 
such gas approaches the 'Compton temperature', $T_{IC}\sim 10^7$K, although
multidimensional hydrodynamic simulations show that due to adiabatic losses the temperature may be smaller,
$\sim 10^{6}\,$K \citep{Dorodnitsyn08a,Dorodnitsyn08b}.
It also follows that mass loss rates of ${\dot M} = 0.1-1\, M_{\odot}/{\rm yr}$
and velocities of $100-1000\,{\rm km\,s^{-1}}$ are attainable, 
and that these properties match those of X-ray warm absorbers.  

Free electrons in the wind scatter 
radiation from the accretion disk and broad line region towards the observer
giving rise to the polarized optical and UV \citep{AntonucciMiller85} spectrum in Seyfert 2 objects. 
Highly ionized ions from this same mirror  gas create rich X-ray spectra of warm absorbers
at inclinations appropriate for Seyfert 1 objects. 
These correspond to photoionized plasma with numerous absorption and emission lines from highly ionized 
ions of Fe, Si, S, O, and Ne in the 0.1 - 10 keV energy range.
These are detected by the $Chandra$ and $XMM-Newton$ satellites,
in addition to features from many of the same ions in the UV seen by e.g. $HST$.
Such spectra occur in a large fraction of low redshift AGNs which 
are bright enough to allow detection \citep{Reynolds97,McKernan07}. 
These lines occur in absorption in Seyfert 1 galaxies, and in reflection spectra in Seyfert 2 galaxies 
\citep{Kinkhabwala02}, corresponding to differing views of the flow and torus.

X-ray polarimetry promises a new opportunity to study AGN winds through their 
X-ray reflection and thus to test AGN unification models. 
Here we report the results of self-consistent 
modeling of the X-ray polarization signature produced by a warm absorber flow.  
We choose a torus model with initial mass 
$M_{\rm tor}=9.3\cdot 10^{5}\,M_{\odot}$, so the maximum Compton 
optical depth of the torus interior, $\tau_{\rm max}=40$, is high enough to 
ensure capture of essentially all X-rays incident on the torus inner surface. 
We calculate the evolution 
of such gas and the fate of the torus, using the techniques
described in \cite{Dorodnitsyn08a,Dorodnitsyn08b}, which includes the dynamics and the effects of photoionization
and photon heating on the thermal and opacity properties in the gas
\citep{KallmanBautista01}.

\begin{figure}[htp]
\includegraphics[width=220pt]{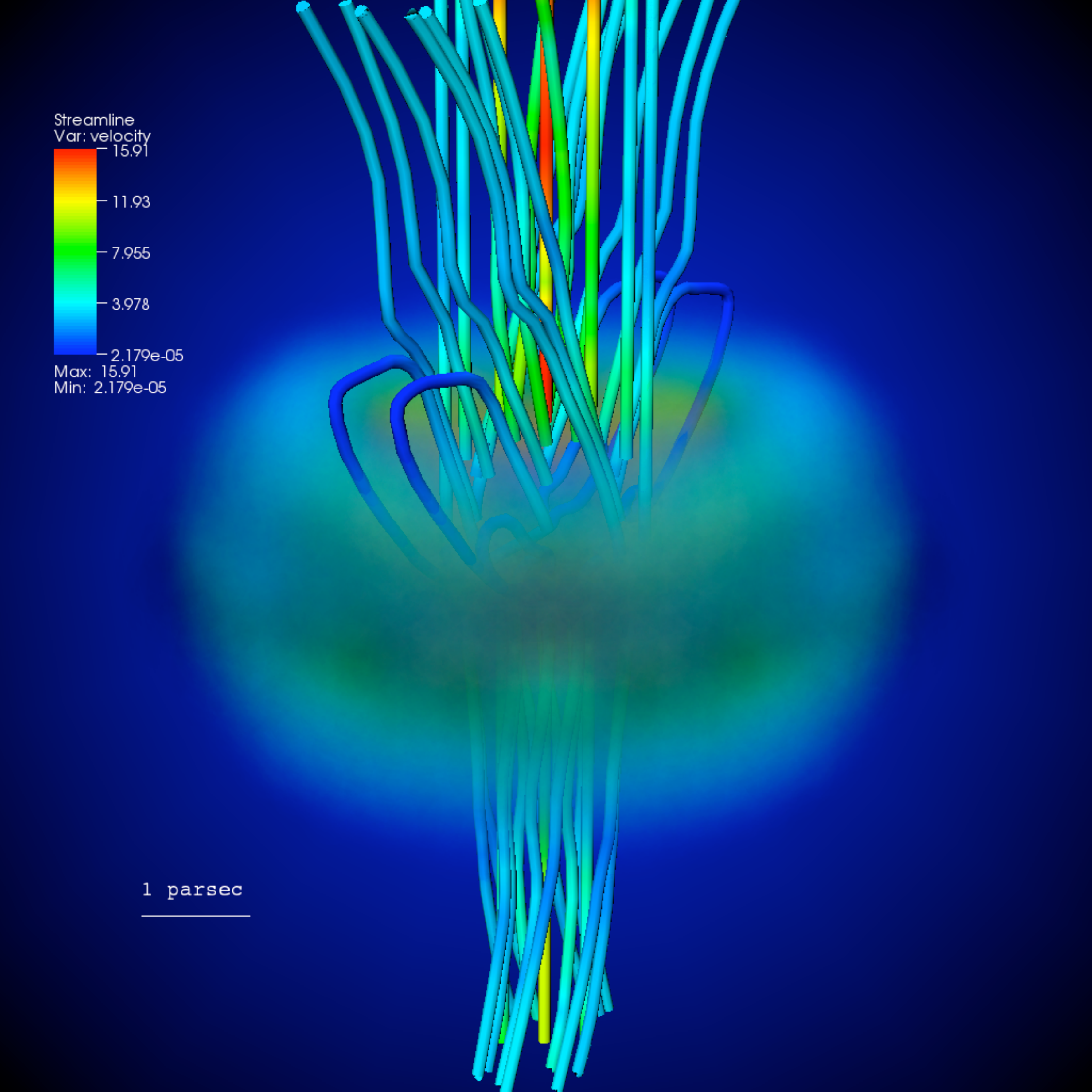}
\caption{The AGN torus, observed at $t=4\,t_{0}$ ($t_{0}$ is the dynamical time) 
from an inclination of $50^{\circ}$. Density is shown 
as the shaded equatorial ring and velocity streamlines are shown as colored tubes emerging from the central hole.
Close to the axis the wind forms a fast bipolar outflow, with too low density and too high ionization to form 
observable warm absorber features.
Favorable conditions for warm absorber viewing exist for  inclinations $40 - 60^{\circ}$.
}\label{torus3D_1}
\end{figure}

\noindent
Gas evaporated by X-ray heating is confined by the torus funnel. 
Our calculations suggest that to some extent, the throat of the torus acts similar to a de Laval nozzle in 
a rocket engine, simultaneously providing the flow with the fuel through evaporation of the material of the
confining walls.
This helps to convert the internal energy of the heated gas to bulk 
kinetic energy of the outflow. Some of the X-ray heated gas is close enough to the black hole to capture significant 
radiation flux and contributes to the polarization signature.
This is enhanced by gas which fails to escape from the potential well, forming 
a returning current, as illustrated in Figure \ref{torus3D_1}.
The opening of the torus funnel is affected by the hydrodynamics of evaporation and
determines the fraction of AGN in which warm absorber spectra can be observed \citep{Dorodnitsyn08b}.

\section{Polarization signature of the X-ray excited flow}
Our calculations consist of three parts. The first part includes time-dependent, 2.5-dimensional hydrodynamic simulations, i.e. time-dependent 2-dimensional simulations with rotation.  The equations of hydrodynamics are 
solved using the ZEUS-2D code \citep{StoneNorman92}, which we have modified to take into account various processes of radiation-matter interaction. The extensive heating occurs on a timescale which is much shorter than the dynamical time, which makes the system of equations stiff, and thus including such processes requires a modification of the ZEUS hydrodynamical scheme \citep{Dorodnitsyn08b}.
Radiation-matter interactions include: Compton and photoionization heating and Compton, radiative recombination, bremsstrahlung and line cooling. The XSTAR \citep{KallmanBautista01} code is used to calculate rates of heating-cooling, and also to calculate tables of line opacities.
These are used in the final stage, when we calculate the emergent spectrum. 
We calculate the transfer of polarized radiation in spectral lines in 
the rapidly moving medium adopting the \cite{Sob60} approximation, 
together with a three-dimensional version of the method of escape probabilities \citep{Castor70,RybickiHummer83}.

The gas evaporated from the surface of the torus by X-ray heating 
is confined within the torus funnel within a vertical range of $z\sim 1-2$ pc. The sonic surface is located at, or very close to the surface
of the funnel (see Fig. 3 of \citet{Dorodnitsyn08b}, where the flow dynamics is described in detail). 
The torus maintains aspect ratio of $R/H\sim 1$ until very late times when it is significantly depleted by the mass loss.
Simulations show that the flow is very highly ionized close to the axis with a temperature approaching Compton temperature, and at higher inclinations as the wind is more dense, self-screening provides favorable conditions for the formation of warm absorber spectra. 
At inclinations where warm absorber gas is observed, the line-of-sight first touches the stripped cool gas of the torus outskirts, and then penetrates through the torus wall. The distribution of the ionization parameter, $\xi$ varies dramatically along the line-of-sight.
At inclinations, $\theta>50^{\circ}-60^{\circ}$, the radiation of the nucleus is heavily obscured. 
Warm absorber spectra are produced within a range of angles $\Delta\theta\simeq 10^{\circ}-15^{\circ}$ at such inclinations. The inferred velocity of the gas decreases as the inclination increases.

The results of hydrodynamic modeling are post-processed with our radiation transfer code. This code makes use of the Sobolev approximation to calculate a three-dimensional transfer of the polarized radiation in approximately $10^{5}$ spectral lines in 0.1 -10 keV range. A precursor of this approach included 3D Sobolev radiation transfer of the un-polarized radiation and is described in \cite{DorodnitsynKallman09}. 
Scattering of the polarized radiation in a resonant spectral line with corresponding calculations of the scattering  phase matrix, was studied by, e.g. \cite{ChandrasekharRadTransfer}, \cite{Hamilton47}, \cite{LeeBlandfordWestern94}.
However, calculations which take into account transfer of polarized light through a 
general 3-dimensional distribution of opacity are numerically demanding.  To make them tractable for the many lines in a real gas, significant 
modification is required both to the methods and the algorithm.  

Following Chandrasekhar, we describe the intensity of the polarized radiation by the Stokes vector:
${\bf I}=(I_{\rm L},I_{\rm R}, U, V)$, where $I_{\rm L}$, $I_{\rm R}$ are the intensity components projected on two arbitrary, orthogonal axes;
$U$ measures the polarization $45^{\circ}$ away from the axis, and $V$ measures circular polarization. The equation, describing the transfer of the polarized  radiation in arbitrary direction $\bf s$, reads, (e.g. \cite{ChandrasekharRadTransfer}): 

\begin{equation}\label{RadiationTransferEq1}
({\bf n}\cdot \nabla) {\bf \IS}= \frac{d\IS}{d{\bf s} }= (\chi_{\rm l}\,\varphi)\, (\Ss-\IS)\mbox{,}
\end{equation}
where $S$ is a Stokes vector source function, $\nu^{\infty}$ is a frequency of the photon, measured in the ''lab'' frame, and 
$\varphi$ is the spectral line profile, and $\chi_{\rm l}$ is the line opacity.
Every resonance with a spectral line, is accompanied by a corresponding change in the radiation field:

\begin{equation}\label{Intens_jump_onEFS}
{\bf I}_{\rm after}={\bf I}_{\rm before} e^{-\tau_{l,i}}+
{\bf S}(1-e^{-\tau_{l,i}})={\bf I}_{\rm dist, i}+{\bf I}_{\rm loc, i}\mbox{,}
\end{equation}
that is, the radiation field consists of the local and distant contributions;  the Sobolev optical depth $\tau_{\rm l}$ \citep{Castor70,RybickiHummer78,DorodnitsynKallman09}
is assumed to be independent of polarization. 
The first term in (\ref{Intens_jump_onEFS}) gives the attenuation of the external radiation and the
${\bf S}(1-e^{-\tau_{l,i}})$ term is the reinforcement of the radiation due to scattering of the radiation from all
directions into the direction towards an observer.

The Sobolev optical depth scales the escape probability, $\beta_{\bf n}$ which is given by
\begin{equation}
\beta_{\bf{n}} = (1-e^{-\tau_{l,i}( {\bf n} )} )/e^{-\tau_{l,i}( {\bf n} ) }
\mbox{.}
\end{equation}
Integrating (\ref{RadiationTransferEq1}) we adopt 
the disconnected approximation \citep{GrachevGrinin,Marti77,RybickiHummer78,RybickiHummer83}; 
that is, we assume that the interaction between resonances along the line-of-sight is happening 
only in the forward direction (towards the observer), and that all other possible interactions are neglected.
Only resonant line transitions are considered, and complete frequency redistribution in the co-moving frame is assumed.

The source function, ${\bf S}$ is calculated from the following \citep{Hummer69}:

\begin{equation}\label{SourceFuntion_def}
{\bf S}(\nu {\bf n} ) = \frac{4\pi}{\varphi} \int\, {\bf R} (\nu' {\bf n}',\nu {\bf n} )\, {\bf {J}}(\nu'{\bf n}) \,
d \nu' d\Omega'\mbox{,}
\end{equation}
where  ${\bf R}(\nu' {\bf n'},\nu {\bf n} )$ is the scattering phase matrix, given by \cite{ChandrasekharRadTransfer}.
This matrix is a generalization of the case 
of a linear combination of the Rayleigh and isotropic phase matrices;
coefficients of the matrix depend on the angular momentum quantum numbers of the initial and final 
states of the transition \citep{Hamilton47}.
The mean intensity, ${\bf J}(\nu)$ depends on the availability of radiation, and thus on a source function, which itself depends on angles through its dependence upon $\beta_{\bf n}$. 

We take approximately into account the depolarizing effect of finite Sobolev optical depth, and
thus avoid overprediction of the polarization fraction.
That is, when the local escape probability, $\beta_{\rm esc}<<1$, the photon suffers 
many uncorrelated resonance scatterings within the resonant region, 
which tend to depolarize and make isotropic the local contribution to the radiation field. 
Solving such a problem in a medium with spatial velocity gradients is better done with a local Monte Carlo simulation. However, for our purposes, that 
would be numerically prohibitive.  
To make the problem tractable,  we adopt a simplifying approach 
in which the source function is represented as a combination of 
the source function derived for an isotropic phase matrix for the local contribution to the radiation field, and 
an anisotropic part for the distant contribution.
The mean intensity ${\bf J}$ consists of two parts: one is the local contribution which depends on the source function, ${\bf S}$, and the other is given by the radiation
coming from the external source (the ''core'').
The idea is to calculate ${\bf J}$, form the angular independent, {\it un-polarized} source function, ${\bf S}^{(0)}$ , and then to substitute it in (\ref{SourceFuntion_def}). Such an approximation captures important properties of Sobolev transfer of the polarized radiation, anticipated also on physical grounds.
Making use of the method of escape probabilities \citep{Castor70}, ${\bf S}$ is found with the further help of the relation:
${\bf J} = {\bf S}^{(0)} (1-<\beta_{\bf n}>_{4\pi}) + <{\bf I}_{\rm c}\beta_{\bf n}>_{\Omega_{\rm src}}$,
where we make use of the notation: $<f>_{\Omega}=(4\pi)^{-1}\int_{\Omega}\,f(\Omega')\,d\Omega'$.  The last term represents the probability for the radiation of the core, $\bf I_{\rm c}$ to reach a given
point, weighted by the probability of penetration.
We assume that the core emits continuum radiation of specific intensity $I_{c}$ which is constant near the frequency of the line. 
For the source function which takes into account the effects of polarization we obtain:
\begin{equation}\label{SourceFunctionPolarized}
{\bf S}^{(1)}({\bf n}) = \int
{\bf R}({\bf n'},{\bf n} ) \left\{ {\bf S}^{(0)}(1-\beta_{\bf{n}'} )+ <{\bf I}_{\rm c}\beta_{\bf{n}'}>_{\Omega_{\rm src}}
\right\} \, d\Omega'\mbox{,}
\end{equation}
where the un-polarized source function, ${\bf S_{0} }$ is found from the relation:

\begin{equation} \label{SourceFunctionUnpolarized}
{\bf S}^{(0)} = \frac{1}{4\pi} \frac{<{\bf I}_{\rm c}\beta_{\bf n}>_{\Omega_{\rm src}} } { \beta_{\rm esc} }\mbox{.}
\end{equation}
In case of isotropic scattering, from (\ref{SourceFunctionPolarized}) one will recover (\ref{SourceFunctionUnpolarized}), i.e. the un-polarized 
source function, which is usually used in Sobolev calculations of the line spectra. 
The other extreme is $\beta_{\bf n'}=1$; all photons arriving from the core are scattered and leave the resonance region without any further interaction, so that no gradient-related effects are important. On the other hand, if $\beta_{\bf n'}<<1$ a given photon is scattered many times within the same local resonant region. The resultant intensity may get significantly depolarized. 

The resultant polarization of the local contribution to the intensity is largely determined by the last scattering event, i.e. by the
${\bf R}( {\bf n'}, {\bf n} )\, {\bf S}^{(0)}$ term. Further details of the method including the geometry are given in \cite{DorodnitsynKallman09} for the non-polarized case
(for the geometry of the problem compare with their Figure 1). Here we formally generalize to the case where four components of the Stokes vector are transfered. We have rigorously tested our code with respect to problems of Sobolev transfer of the polarized flux in individual lines and the distribution of lines.
Prescribing different distributions of the velocity and density we are able to compare transfer of the individual polarized modes with 
results of un-polarized transfer calculations.

\begin{figure}[htp]
\includegraphics[width=320pt]{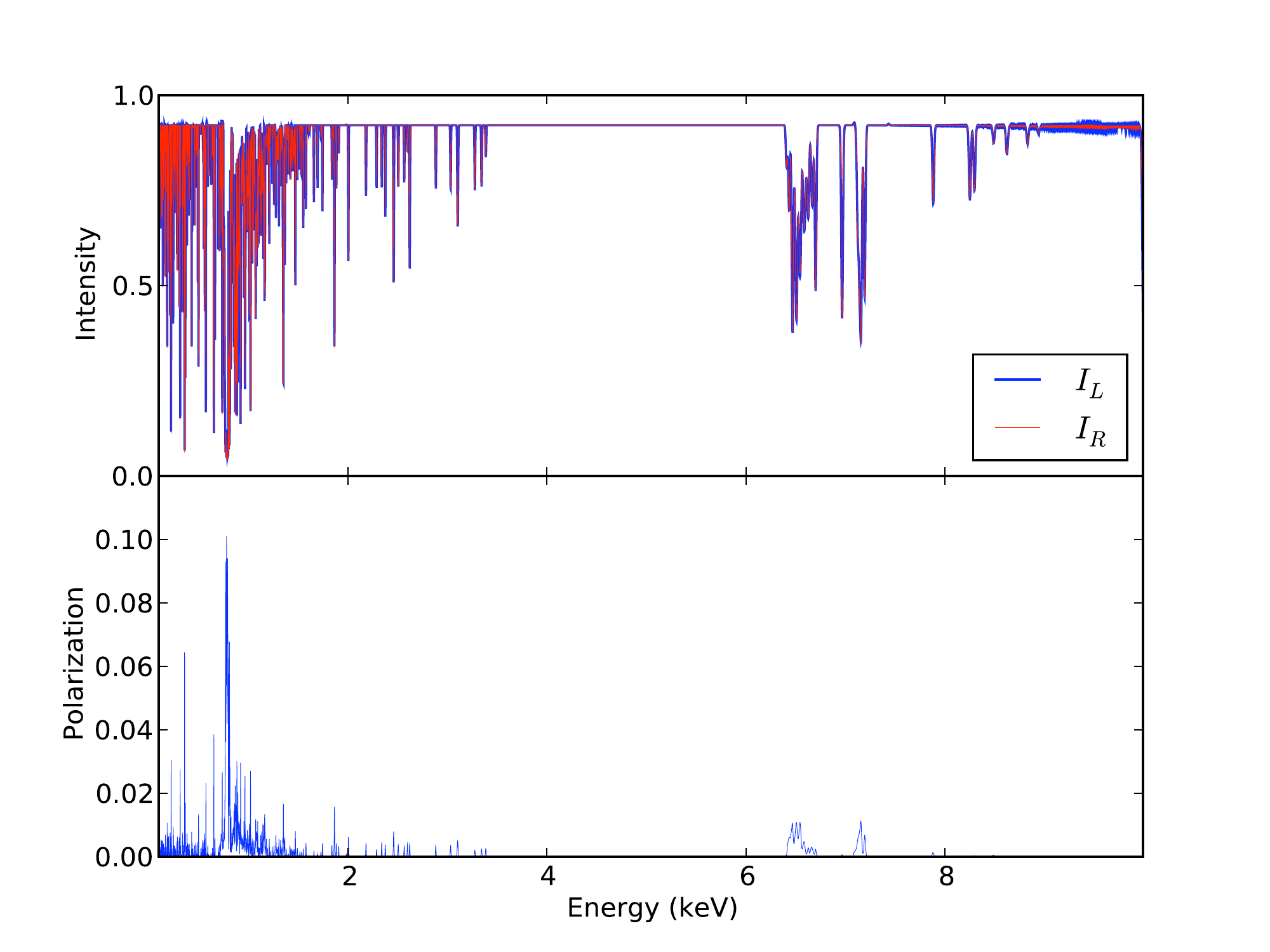}
\caption{Warm absorber spectra viewed at inclination $50^{\circ}$ in two orthogonal linear polarization modes, $I_{\rm L}$,$I_{\rm R}$ (up). Polarization fraction, $P$ (bottom).
}\label{FigSpectr_50}
\end{figure}

\begin{figure}[htp]
\includegraphics[width=320pt]{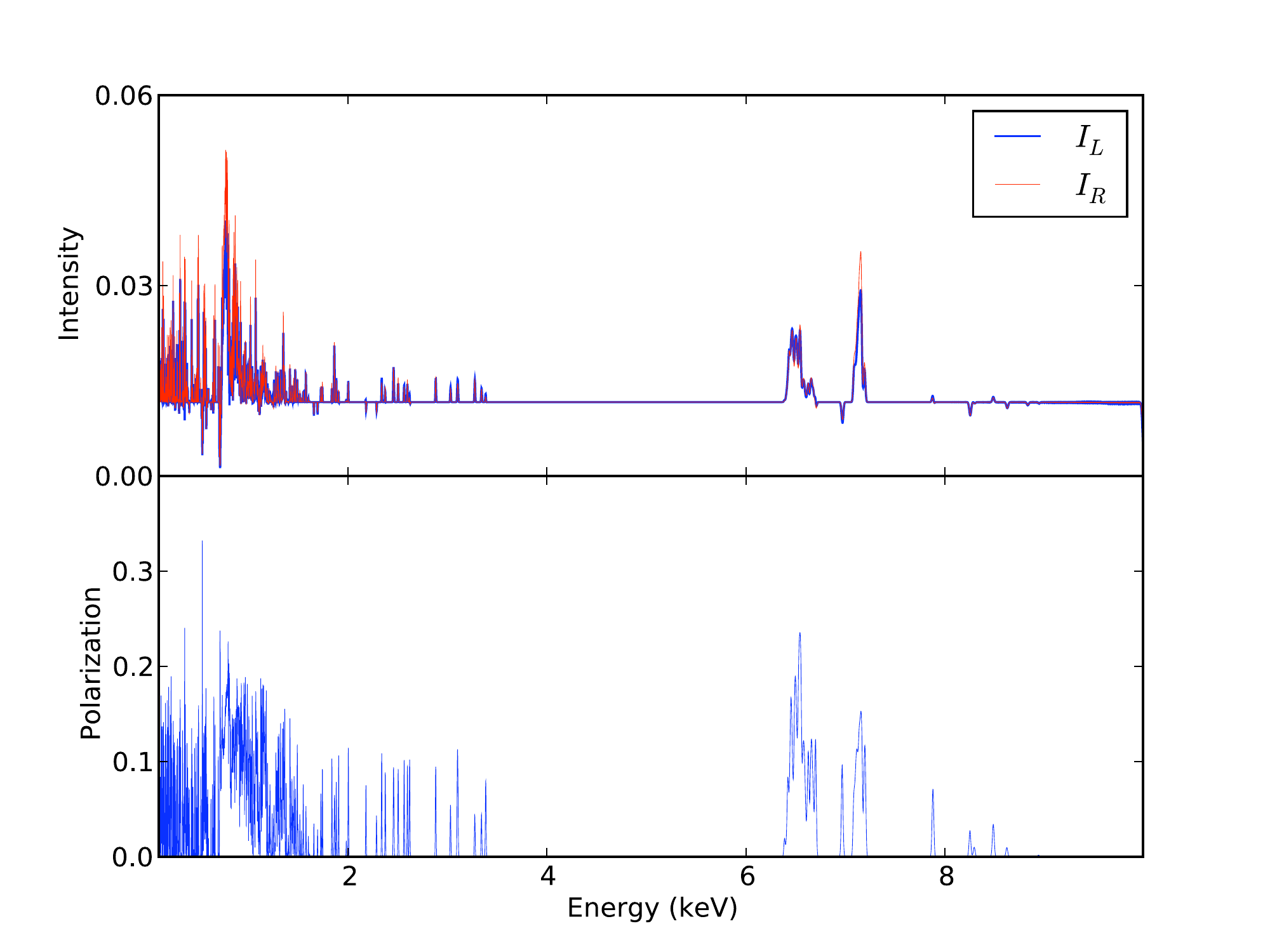}
\caption{Same model as in Figure \ref{FigSpectr_50} but at inclination $70^{\circ}$.
}\label{FigSpectr_70}
\end{figure}

\section{Results}
A notable feature of most warm absorber spectra is the fact few of the lines appear truly black in their 
cores.  While this could be due to finite resolution or finite column of the gas, in our simulations 
this is because of scattering into the line core.  Such scattering typically occurs at angles close to 90$^o$, and 
so the residual flux in the lines is highly polarized.  This is similar to the polarization of the (rest frame) UV lines 
observed from broad absorption line quasars   \citep{Ogle99}. 
Figure \ref{FigSpectr_50} shows spectra and the corresponding polarization fraction, 
$P$ at inclination $\theta=50^{\circ}$ after the wind simulation has evolved
through four dynamical times  
(the dynamical time is given by: $t_{0}=1.5\times 10^{4}\,r_{\rm pc}\, (M_{\rm BH}/M^{6}_{\odot}) ^{-1/2}\,(\rm yr)$,
where $M_{\rm BH}/M^{6}_{\odot}$ is the BH mass in units of $10^6 M_{\odot}$ and $r_{\rm pc}$ is the radius of the torus in pc)
Each synthetic spectrum is convolved with a Gaussian with width and centroid related by $\sigma_{E}=10^{-3}E$.
In the 0.1-10 keV range, the most prominent feature of the spectrum is  
an absorption trough (Figure \ref{FigSpectr_50}, upper panel) and a corresponding 
maximum in polarization at $\sim 14.9 -17.2\As,\ (0.72 - 0.83$ keV).
This is due to H-, and He-like oxygen and moderately ionized iron, ${\rm Fe\,{XVII} 
- Fe\,{XIV}}$. The polarization maximum has $P_{\rm max}=10\%$, and polarization 
at FWHM=0.05 keV about 2\%.  This is produced at a distance $\simeq$2.1 pc from the BH, 
where the  radial column density rises from $N_{\rm col}=1.5\cdot 10^{22}$ to $N_{\rm col}=3.8\cdot 
10^{22}$, and the ionization parameter, $\xi$  \citep{Tarter69} drops from 
$6.4\cdot 10^{3}$ to $500-10$ within a region $2.8-4.4$ pc.  This results in an ideal 
situation for scattering in the flow, since further away the plasma is 
again over-ionized: $\xi\sim 10^{4}$.  Polarization of lines in the 0.1-2 
keV band is generally $\leq$1\%, but for the strongest lines the polarization can 
reach 2 -- 3\% in the O VIII L$\alpha$ and Ne IX He $\alpha$ lines; and 
$\simeq$10\% in the case of the  features at $\sim 0.75$ keV.

At high inclinations, $\theta\gtrsim 60^{\circ}$, most lines appear in emission, and 
the continuum flux is $\sim 10^{-2}$ of that at low inclinations. 
Figure \ref{FigSpectr_70} shows the spectrum and corresponding polarization fraction, 
$P$ at inclination $\theta=70^{\circ}$. 
One of the insights provided by multi-dimensional transfer calculations is the importance of blending of
multiple lines, i.e. when several lines collectively appear as a single strong feature.
Blending of multiple lines from Fe XXV and Fe XXVI forms
such a spectral feature at $E_{0}\sim 6.5$ keV, at the inclination in figure \ref{FigSpectr_70},  resembling 
a P-Cygni line. However, this shape is not due to the effect of  
the moving wind, rather it is an interplay between geometric effects and ionization 
balance.  That is, due to the attenuation of the continuum at high inclinations,  
when the continuum intensity in the line emitting region falls below a threshold, 
$I_{\rm c}\lesssim S_{\rm l}(1-e^{\tau})$, where $S_{\rm l}$ is the source 
function and $\tau_{\rm l}$ is the optical depth, then the ensemble of lines corresponding 
to lower ionized gas pops up above the continuum, while the more highly ionized 
gas is still seen in absorption. 
Other examples of such features occur at lower inclinations, such as at
$\theta\simeq 45^{\circ}$, corresponding to the  ${\rm O\,{VII}}$ edge at 
16.8$\As$ (739 eV), together with lines from the L shell of Fe.

\begin{figure}[htp]
\includegraphics[width=350pt]{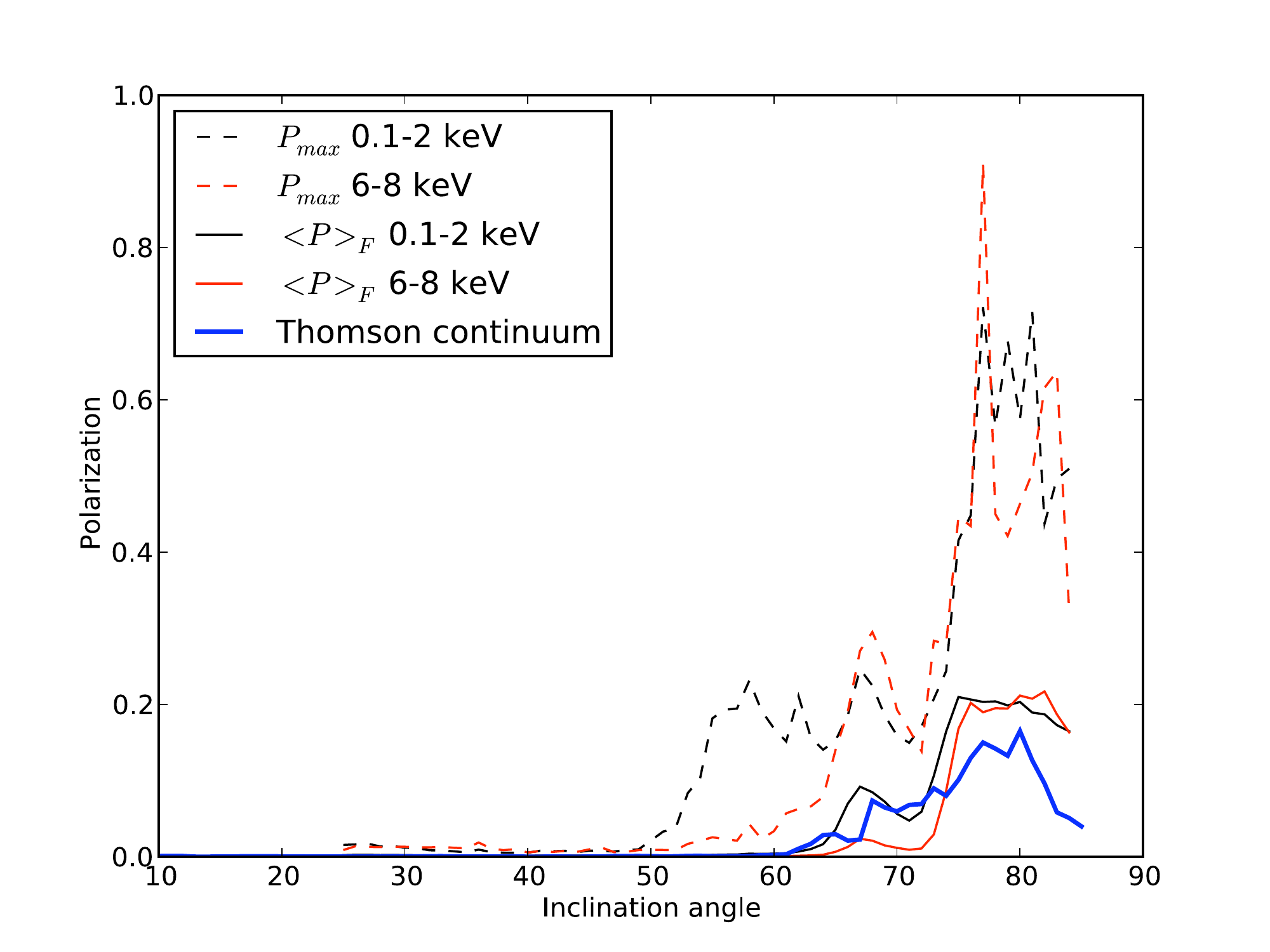}
\caption{Distribution of the polarization fraction with inclination angle. Curves: dashed: maximum polarization fraction within the energy range;
solid: thin: flux mean polarization fraction within the energy range; thick: Thomson scattering continuum
}
\label{Fig_POL}
\end{figure}

Figure \ref{Fig_POL} shows the distribution of polarization fraction
versus inclination angle. We plot the maximum polarization, 
$P_{\rm max}$ ($P$ convolved with a  Gaussian with $\sigma_{E}=(15)^{-1}E$),  
and the flux mean $<P>_{\rm F}$ in the 0.1-2 and 2-10 keV bands.
This shows that the maximum polarization is $<P>_{\rm max}\simeq 15-20\%$ over a wide range of angles.
The various peaks and troughs in this plot are reflections of the density structure in the flow,
including wedges of gas which are stripped off the torus.
The flux mean polarization has a maximum near $\theta=67^{\circ}$, where 
$<P>_{\rm F}(0.1-2) \,{\rm keV}\lesssim 10\%$ and $<P>_{\rm F}(2-10)\, {\rm keV}\lesssim 2\%$.  
At  $\theta\gtrsim 75^{\circ}$ the flux mean polarization in the 0.1-10 keV range is $<P>_{\rm F}\lesssim 20\%$  
and the maximum is  $P_{\rm max} \simeq 70\%$.  
At such inclinations the torus is Compton thick along the line of sight and the continuum flux is very low, 
$\leq 1\%$ of the unattenuated value.

Scattering by free electrons from optically thin plasma can produce a continuum polarization fraction which is comparable 
to that in the lines \citep{Bianchi09}. Optically thin, single scattering approximation provides an upper limit 
for the polarization.
This approach takes into account the geometry of the problem together with optical depth effects 
and predicts both the polarized flux and the fractional polarization is less than those produced by lines.  This is 
shown as the labeled curve in Figure \ref{Fig_POL}). At those viewing angles where the Thomson continuum 
has significant polarization, the flux is $\sim (10^{-3}-10^{-4})F_{\rm c}$, where  $F_{\rm c}$ is the flux of the core.

Our results show that models in which the obscuring torus is the source of an X-ray excited flow 
generate potentially observable polarization signatures due to 
significant scattering of  X-rays by atomic features in the flow.
Thomson electron scattering continuum at $\theta\lesssim 60^{\circ}$ has polarization $<1\%$; it is increasing to a maximum of $17\%$ at $\theta\simeq 80^{\circ}$, and then
decreases at larger $\theta$.
Polarization from lines is greater and has the following properties:
at very low inclination, $\theta\lesssim 45^{\circ}$, corresponding to prototypical Seyfert 1 galaxies, 
the linear polarization fraction is negligible averaged 
over the spectrum, but  most absorption lines have some residual, highly polarized flux;
the maximum polarization fraction in strong lines is $1 - 10 \%$. 
At higher inclinations, $45^{\circ}\lesssim\theta\lesssim75^{\circ}$ 
in the  0.1-2 keV range the mean polarization is 10\%, and the maximum polarization is 20-30\%;
in the  6-8 keV range the mean polarization is 2-10\%, and the maximum polarization 
20-30\%.  At still higher inclinations $\theta\gtrsim 70-80^{\circ}$
our models are very Compton thick.  Scattering in X-ray lines above the torus funnel
produces very strongly polarized flux
with mean polarization fraction more than 20\% and maximum polarization 60-70\%.
Although these predictions are specific to our dynamical calculations, the general behavior will hold 
for any scenario involving an outflow which is shadowed by a torus with 
aspect ratio near unity.  Observations which fail to show such high polarization 
would call into question current ideas about the torus location, shape and thickness.

\hbox{}
This research was supported by an appointment at the NASA Goddard Space Flight Center, administered by CRESST/UMD through a contract with NASA, and by grants from the NASA Astrophysics Theory Program 05-ATP05-18.

\end{document}